\documentclass[11pt]{article}
\usepackage[english]{babel}
\usepackage{graphicx}
\usepackage{mathrsfs}
\usepackage{amsmath}
\usepackage{amssymb}

\long\def\symbolfootnote[#1]#2{\begingroup\def\thefootnote{\fnsymbol{footnote}}
\footnote[#1]{#2}\endgroup}
\begin{document}
\title{\Large\bf Intertwiner dynamics in the flipped vertex}

\author{Emanuele Alesci {\it ${}^{1ab}$}, Eugenio Bianchi {\it ${}^{2ac}$}, Elena Magliaro {\it ${}^{3ad}$}, Claudio Perini {\it ${}^{4ae}$} \\
[1mm]
\small\it ${}^a$Centre de Physique Th\'eorique de Luminy\footnote{Unit\'e mixte de recherche (UMR 6207) du CNRS et des Universit\'es de Provence (Aix-Marseille I), de la M\'editerran\'ee (Aix-Marseille II) et du Sud (Toulon-Var); laboratoire affili\'e \`a la FRUMAM (FR 2291).}\;, Case 907, F-13288 Marseille, EU\\
\small\it ${}^b$Laboratoire de Physique, ENS Lyon, CNRS UMR 5672, F-69007 Lyon, EU\\
\small\it ${}^c$Scuola Normale Superiore, I-56126 Pisa, EU\\
\small\it ${}^d$Dipartimento di Fisica, Universit\`a degli Studi Roma Tre, I-00146 Roma, EU\\
\small\it ${}^e$Dipartimento di Matematica, Universit\`a degli Studi Roma Tre, I-00146 Roma, EU
}

\date{\small\today} \maketitle
\symbolfootnote[0]{e-mail: ${}^{1}$alesci@fis.uniroma3.it, ${}^{2}$e.bianchi@sns.it, ${}^{3}$elena.magliaro@gmail.com, ${}^{4}$claude.perin@libero.it}
\begin{abstract}
We continue the semiclassical analysis, started in a previous paper, of the intertwiner sector of the flipped vertex spinfoam model. We use independently both a semi-analytical and a purely numerical approach, finding the correct behavior of wave packet propagation and physical expectation values. In the end, we show preliminary results about correlation functions.
\end{abstract}
\vskip1cm

\section{Introduction}
In \cite{EPR}\cite{Engle:2007qf}\cite{Engle:2007wy}\cite{Pereira:2007nh} a new spinfoam vertex, now called the EPR or ELPR$\gamma$ vertex, was introduced. This model, as well as other new models \cite{Alexandrov:2007pq}\cite{Freidel:2007py}\cite{Livine:2007vk}, was born essentially in order to correct two problems of the Barret-Crane vertex \cite{BC}: the wrong intertwiner (non) dependence and the mismatching with the Loop Quantum Gravity \cite{LQG}\cite{Thiemann}\cite{AshLew} boundary states. As shown in \cite{Alesci1}, the BC model has a good semiclassical behavior in the spin sector, but not in the intertwiner sector; actually in the BC theory the intertwiner degrees of freedom are dynamically frozen. In particular, the diagonal components of the graviton propagator have the right scaling with distance \cite{Modesto:2005sj}\cite{Rovelli:2005yj}\cite{Bianchi:2006uf}\cite{LivSpez2006}\cite{NumericalPropagator}\cite{Bianchi:2007vf}, but the nondiagonal (the ones depending on the intertwiners) have the wrong scaling \cite{Alesci1}\cite{Alesci2}. An important open issue is to check if the new spinfoam models correct those pathologies. A new technique to test the spinfoam dynamics, which is complementary to the calculation of $n$-point functions, was introduced in \cite{numerical}, and a partial answer was given. This technique is the propagation of semiclassical wavepackets: as in ordinary quantum mechanics, if the theory has the correct semiclassical limit, then semiclassical wavepackets must follow the trajectory predicted by classical equations of motion. In \cite{numerical}, the wavepacket propagation in the intertwiner sector was studied numerically, finding a surprisingly good semiclassical behavior. In brief, the authors considered the solution of discretized Einstein equations given by a single flat 4-simplex with boundary constituted by five regular tetrahedra. In the dual LQG picture this boundary is represented by a pentagonal 4-valent spinnetwork, labeled by ten spins and five intertwiners; but in order to have a semiclassical state one has to construct some (infinite) linear combination of spinnetworks of this kind. It is known from \cite{Rovelli:2006fw} that basis 4-valent intertwiners with some choice of pairing can be superposed with gaussian weight  to be able to catch the classical geometry: since in a quantum tetrahedron the angles do not commute, one has to consider semiclassical superpositions to peak all angles on the same classical value. In \cite{numerical} the authors chose an initial state formed by four coherent intertwiners at four nodes, and they made the drastic approximation of taking the ten spins fixed to be equal to some $j_0$. Then they calculated numerically its evolution, here called 4-to-1-evolution, that is its contraction with the propagation kernel of the flipped vertex spinfoam model. Classical Einstein equations impose the final state to be a coherent intertwiner with the same geometrical properties (mean and phase). They found good indications but, due to the very low $j_0$'s, they couldn't conclude much about the analytical properties of the evolved state. Here we conjecture the general behavior of the evolution at high $j_0$'s which is very well supported numerically. In fact, as we shall argue, the propagation is perfectly ``rigid": four gaussian wavepackets evolve into one gaussian wavepacket with the same parameters, except for a flip in the phase. The phase of the evolved phase, and in particular its flipping, will have a major role when considering physical expectation values. The support to our conjecture is made in two independent ways: the first is semi-analytical and is based on a numerical result on the $15j$-symbol viewed as a propagation kernel, and the asymptotic properties of the fusion coefficients already studied in \cite{fusion}; the second is purely numerical. The first has the advantage of giving a nice picture of the dynamics in terms of wavepackets evolving separately in the left and right $SO(4)$ sectors, and it also pave the way for the completely analytical approach (to do this, one should have an asymptotic formula for the $15j$-symbol). We also explore the possibility of propagating three coherent intertwiners into two (we will refer to as the 3-to-2-evolution), finding similar results. Then we present the results from another point of view, namely as intertwiner physical expectation values, finding that these are asymptotically the predicted ones. Though we use the drastic approximation of fixing all spins, we regard our results as a strong indication that the EPR model has the good semiclassical limit. In the end we present the numerical calculation of the intertwiner correlation function, finding a scaling law which is not the Newtonian one; we believe that this is due to the drastic approximation of freezing the spin variables in the boundary state, and not to pathologies of the model.
\section{The flipped vertex fusion coefficients}
\begin{figure}[h]
\centering
\includegraphics[width=3.5cm]{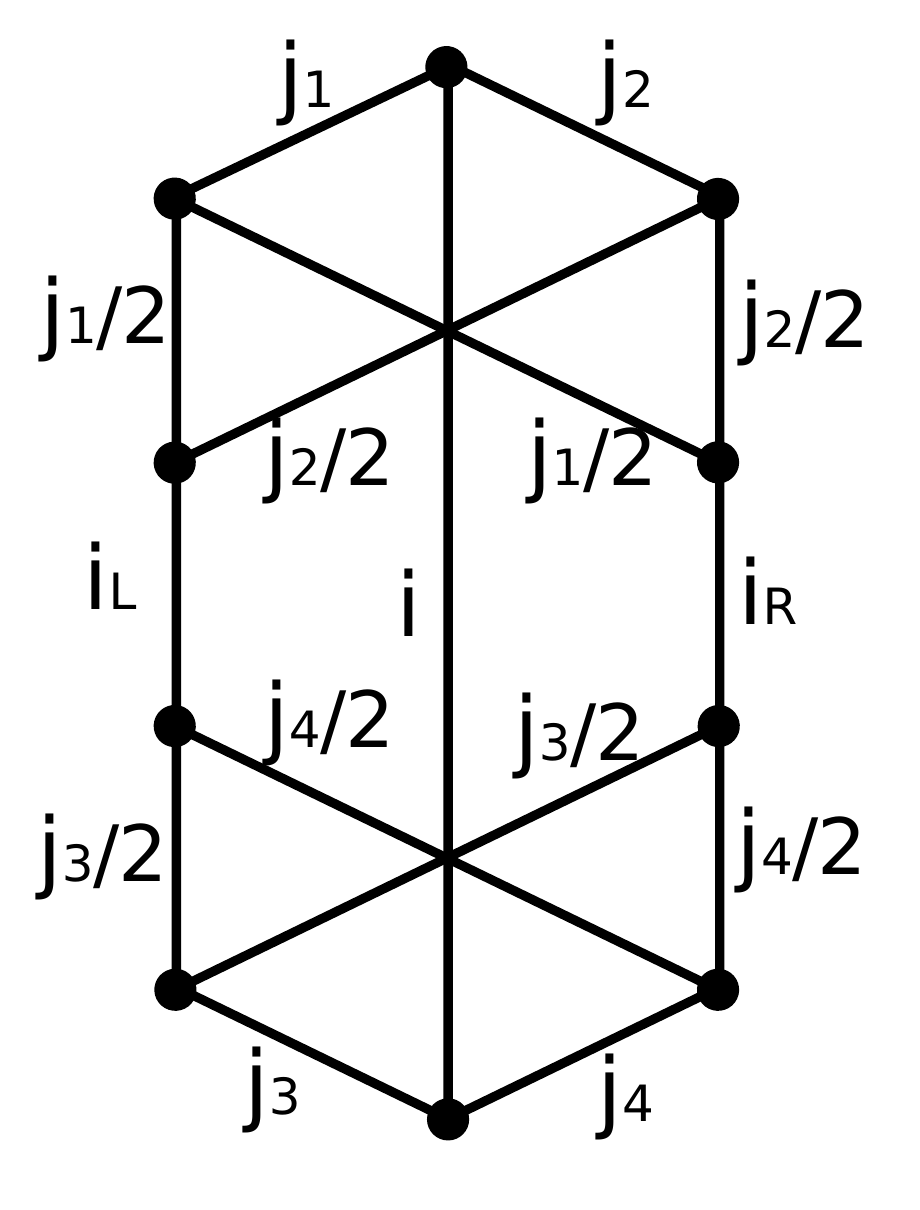}
\caption{the flipped vertex branching function}
\label{figf}
\end{figure}

One of the key ingredients of the EPR dynamics is the branching function or fusion coefficients  $f^i_{i_L,i_R}$, which are the coefficients of a linear map from $SU(2)$ to $SO(4)$ intertwiner vector space appearing in the definition of the vertex. In the case of $\gamma=0$, called flipped vertex, the fusion coefficients are defined by the evaluation\footnote{The evaluation of a 3-valent spinnetwork is defined as the contraction (sum over magnetic numbers) of $3j$-symbols, one for each node, according to the pattern given by the graph.} of the spinnetwork in Fig. \ref{figf}.  
They are symmetric under exchanging of $i_L$ and $i_R$ and non-vanishing only if the relation $|i_L-i_R|\leq i\leq i_L + i_R$ holds. 
In \cite{fusion} the  fusion coefficients were studied and an analytic asymptotic formula was given; thanks to this formula, we found the asymptotical action of the fusion coefficients on a semiclassical intertwiner. We resume briefly some properties. Consider a semiclassical regular tetrahedron, whose faces have (large) area $j_0$; in LQG the tetrahedron, in particular its internal angles, is described by an $SO(3)$ semiclassical intertwiner, defined as
\begin{align}
\label{semiclnode}
|i_0\rangle_{\text{SC}}\equiv\sum_i \psi(i)|i\rangle,
\end{align}
where
\begin{align}
\label{psi}
\psi(i)\equiv C(j_0) \exp\left(-{\textstyle\frac{3}{4 j_0}}(i-i_0)^2+\iota{\textstyle\frac{\pi}{2}}i\right),
\end{align}
$|i\rangle$ is a basis intertwiner between four $SO(3)$ irreps labeled by $j_0$, $C(j_0)$ is a normalization constant and $i_0\equiv 2 j_0/\sqrt3$ ($i$ labels the virtual spin in some pairing). For simplicity, we will refer to \eqref{semiclnode}, or to its components $\psi(i)$, as ``semiclassical intertwiner". We denote with $\iota$ the imaginary unit, in order to avoid confusion with the intertwiner labels. The action of $f^i_{i_L,i_R}$ (viewed as a map between intertwiner spaces) on a semiclassical intertwiner is given by
\begin{align}
\label{gdef}
    g(i_L,i_R)\equiv \sum_i \text{dim}(i) f^i_{i_L,i_R}\psi(i).
\end{align}
We showed that, for large $j_0$'s
\begin{align}\label{gfactorized}
g(i_L,i_R)\simeq C\exp\Big(\!\!-{\textstyle\frac{3}{2 j_0}}(i_L-{\textstyle\frac{i_0}{2}})^2-{\textstyle\frac{3}{2 j_0}}(i_R-{\textstyle\frac{i_0}{2}})^2+\iota{\textstyle\frac{\pi}{2}}(i_L+i_R)\Big),
\end{align}
where $C$ is an irrelevant normalization factor not depending on $i_L$ and $i_R$ at leading order in $1/j_0$ powers. Hence, asymptotically, the function $g$ factorizes into left and right parts; we indicate them, with abuse of notation, $g(i_L)$ and $g(i_R)$. The values of $g(i_L,i_R)$ are the components of an $SO(4)\simeq SU(2)\times SU(2)$ intertwiner in the basis $|i_L,\,i_R\rangle$, which in the next we will call $SO(4)$ semiclassical intertwiner. Also the converse holds: the asymptotical action of the fusion coefficients on an $SO(4)$ semiclassical intertwiner is an $SO(3)$ semiclassical intertwiner, i.e.
\begin{align}
\label{converse}
  \sum_{i_L,i_R} \text{dim}(i_L)  \text{dim}(i_R)\,f^i_{i_L,i_R}\,g(i_L,i_R)\simeq\psi(i).
\end{align}
\section{Propagation of intertwiner wavepackets}
\begin{figure}
\centering 
\parbox{7.5cm}{\includegraphics[width=7cm]{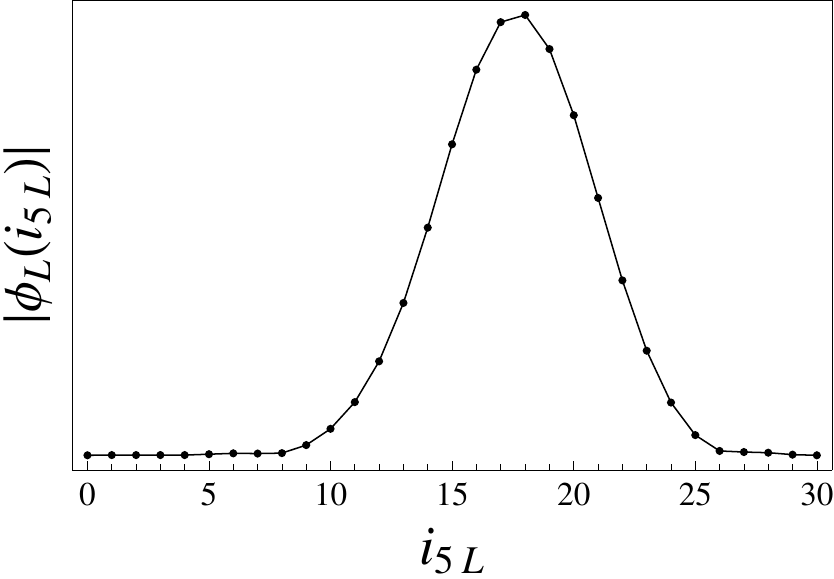}}
\qquad 
\parbox{7.5cm}{\includegraphics[width=7cm]{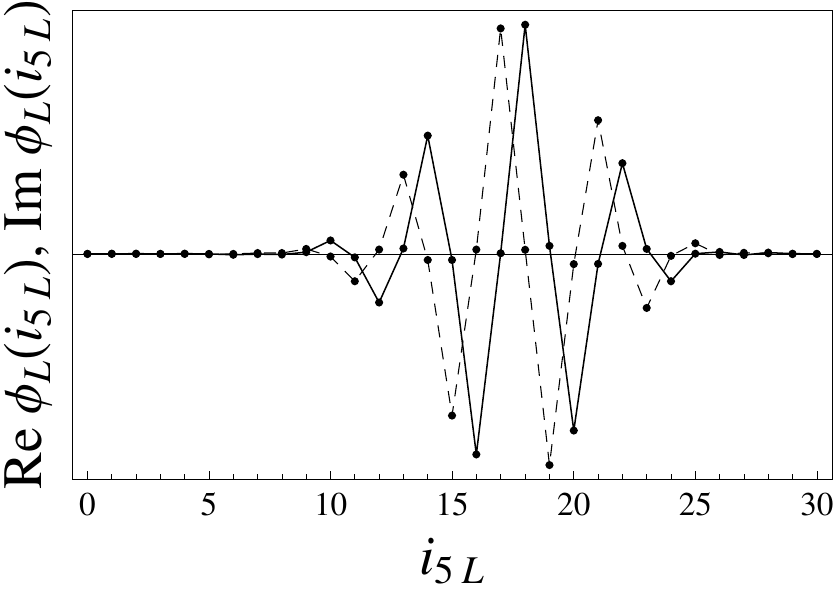}}
\caption{On the left: modulus of the evolved state for the 4-to-1 propagation performed by one $15j$ ($j_0=30$). On the right: its real and imaginary (dashed) part.}
\label{fig15jprop}
\end{figure}

The property \eqref{gfactorized} gives a new picture of the dynamics in the semiclassical regime. Consider a boundary LQG state supported on the dual boundary of a 4-simplex (this state is labeled by ten spins and five intertwiners), and consider the simple case with all spins fixed to be equal to some $j_0$, while the intertwiners are the semiclassical ones, namely:
\begin{equation}
\label{boundary}
\Psi(\{j_n\},\{i_m\})\propto\prod_{n=1}^{10}\delta_{j_n,j_0}\prod_{m=1}^5 \psi(i_m)
\end{equation}
where $\psi$ is the Gaussian \eqref{psi}. This state is the limiting case of a state peaked on the intrinsic and extrinsic geometry of the boundary of a regular classical 4-simplex as it becomes sharp in the spin variables \cite{numerical}. We expect that, if four semiclassical intertwiners are given as initial state, then their evolution under the propagation kernel is again a semiclassical intertwiner (for an introduction and motivations to this idea see \cite{numerical}), the state \eqref{boundary} being peaked on a boundary solution of Einstein equations. Actually this propagation property should be true for a more correct boundary state in which the approximation of freezing the spins is not taken; nevertheless one can study the evolution and see if a positive result is obtained. If this is the case, we are strongly encouraged to believe that the same property holds in the general case. Coming back to details, define the 4-to-1-evolution as
\begin{equation}
\label{4to1}
\phi(i_5)\equiv\sum_{i_1\ldots i_4}\text{dim}(i_{1})\ldots\text{dim}(i_{4})\,W\big(i_1\ldots i_5\big)\, \psi(i_1)\ldots \psi(i_4).
\end{equation}
Here
\begin{align}
\label{flippedvertex}
W(i_1\ldots i_5)\equiv\!\!\!\!\!\!\sum_{\{i_{nL}\}\{i_{nR}\}}\!\left[\prod_{n=1}^5\text{dim}(i_{nL})\,\text{dim}(i_{nR})\right]\,
15j\!\big(i_{1L},\ldots,i_{5L}\big)
15j\!\big(i_{1R},\ldots,i_{5R}\big)\prod_{n=1}^5 f^{i_n}_{i_{nL,nR}}
\end{align}
is the EPR flipped vertex amplitude evaluated in the homogeneous spin configuration (the ten spins are all equal to some fixed $j_0$), and we have omitted the dependence of the $15j$-symbol on the spins. The factor $f^{i_1}_{i_{1L},i_{1R}}\ldots f^{i_4}_{i_{4L},i_{4R}}$ of \eqref{flippedvertex} is contracted in \eqref{4to1} with four initial packets (making the sum over $i_1\ldots i_4$). By \eqref{gfactorized}, for large $j_0$'s this contraction gives four $SO(4)$ semiclassical intertwiners, so the evolved state \eqref{4to1} becomes
\begin{align}\nonumber
\phi(i_5)\simeq\!\!\sum_{i_{5L},i_{5R}}\!\!\text{dim}\,i_{5L}\,\text{dim}\,i_{5R}
&\left[\sum_{i_{1L}\ldots i_{4L}}\!\!\!\!15j\!\big(i_{1L},\ldots,i_{5L}\big) g(i_{1L})\ldots g(i_{4L})\right]\times\\\label{4to1splitted}
\times&\left[\sum_{i_{1R}\ldots i_{4R}}\!\!\!\!15j\!\big(i_{1R},\ldots,i_{5R}\big) g(i_{1R})\ldots g(i_{4R})\right] f^i_{i_{5L},i_{5R}}.
\end{align}
We can see in the last expression the action of two $15j$'s separately on the left and right part (the expressions in square brackets). Those actions are interpreted as independent 4-to-1-evolutions in the left and right sectors, namely the evolution of the left an right part of four SO(4) semiclassical intertwiners, where the dynamical vertex is the $15j$-symbol. By numerical investigations (Fig. \ref{fig15jprop}), it turns out that the final state of the right (left) partial evolution is the right (left) part of an SO(4) semiclassical intertwiner, with the phase flipped as compared to the incoming packets. For example, for the left part:
\begin{equation}
\label{15j4to1}
\phi_L(i_{5L})\equiv\!\!\sum_{i_{1L}\ldots i_{4L}} \text{dim}(i_{1L})\ldots\text{dim}(i_{4L})\,15j\big(i_{1L}\ldots i_{5L}\big) \,g(i_{1L})\ldots g_L(i_{4L})\simeq\overline{g(i_{L5})}.
\end{equation}
Then \eqref{4to1splitted} becomes
\begin{equation}
\phi(i_5)\simeq\sum_{i_{5L},i_{5R}}\!\!\text{dim}\,i_{5L}\,\text{dim}\,i_{5R}\,\,\overline{g(i_{5L})}\,\overline{g(i_{5R})}\,f^{i_5}_{i_{5L},i_{5R}}.
\end{equation}
The last expression is the contraction between the fusion coefficients and an $SO(4)$ semiclassical intertwiner. By \eqref{converse}, this gives an $SO(3)$ semiclassical intertwiner:
\begin{equation}\label{phieqpsi}
\phi(i_5)\simeq\overline{\psi(i_5)}.
\end{equation}
While in \cite{numerical} we expected only a conservation of mean values, and possibly a spread of wave packets, the precedent argument shows that the gaussian shape is conserved, together with its mean value and width, while the phase is flipped. The 3-to-2-evolution is defined similarly to the 4-to-1 case, as the contraction between the flipped vertex and three initial semiclassical intertwiners. Numerical results about this type of evolution are discussed in section \ref{numericalsection}.
\section{Physical expectation values}
In this section we want to present the precedent results from another perspective, as results about expectation values of observables. By construction, the boundary state \eqref{boundary} is peaked {\it kinematically} on a semiclassical geometry. This should be also true in a {\it dynamical} sense, as it is peaked on a solution of Einstein equations. So consider the physical expectation value of an intertwiner on this boundary state:
\begin{equation}
\label{imedio}
\langle i_1\rangle\equiv\frac{\sum_{\{j\},\{i\}} W(\{j\},\{i\})\,i_1\,\Psi(\{j\},\{i\})}
{\sum_{\{j\},\{i\}}W(\{j\},\{i\})\Psi(\{j\},\{i\})}.
\end{equation}
We expect this quantity to be equal to $i_0$ for large $j_0$'s, if the dynamics has the correct semiclassical limit. Analogously, we can consider the expectation value of two intertwiners:
\begin{equation}
\label{iimedio}
\langle i_1\,i_2\rangle\equiv\frac{\sum_{\{j\},\{i\}} W(\{j\},\{i\})\,i_1\,i_2\,\Psi(\{j\},\{i\})}
{\sum_{\{j\},\{i\}}W(\{j\},\{i\})\Psi(\{j\},\{i\})};
\end{equation}
the last expression should be asymptotically equal to $i_0^2$. The results about wavepacket propagation give full information about the previous physical expectation values. In fact, \eqref{imedio} can be viewed as the contraction between the evolved state and a semiclassical boundary intertwiner with one insertion, so
\begin{align}
\label{}
\langle i_1\rangle&=\frac{\sum_{i_1}\text{dim}\,i_1\,\phi(i_1) \,i_1\,\psi(i_1)}
{\sum_{i_1}\text{dim}\,i_1\,\phi(i_1)\psi(i_1)}\simeq \frac{\sum_{i_1} \overline{\psi(i_1)} \,i_1\,\psi(i_1)}
{\sum_{i_1}\overline{\psi(i_1)}\psi(i_1)}=i_0,
\end{align}
where we used \eqref{phieqpsi}; what we have found is that dynamical and kinematical mean (asymptotically) coincide. We stress that, not only the peakiness of the evolved state is required in order to have the right expectation value, but also the phase of the evolved state must cancel exactly the phase of the initial intertwiner. The same properties (peakiness and right phase) hold for the 3-to-2-propagation (see numerical results in the next section), and the expectation value of two intertwiners turns out to be the correct one, i.e. $i_0^2$.
\section{Numerical analysis}\label{numericalsection}
We wrote a numerical algorithm performing the 4-to-1 and 3-to-2 evolutions, and calculating the physical expectation values \eqref{imedio}\eqref{iimedio}; the algorithm computes very big sums serially with a method similar to the one in \cite{efficient}\cite{Igor}. The results are shown in the figures.
\begin{figure}[h]
\centering 
\parbox{7.5cm}{\includegraphics[width=7cm]{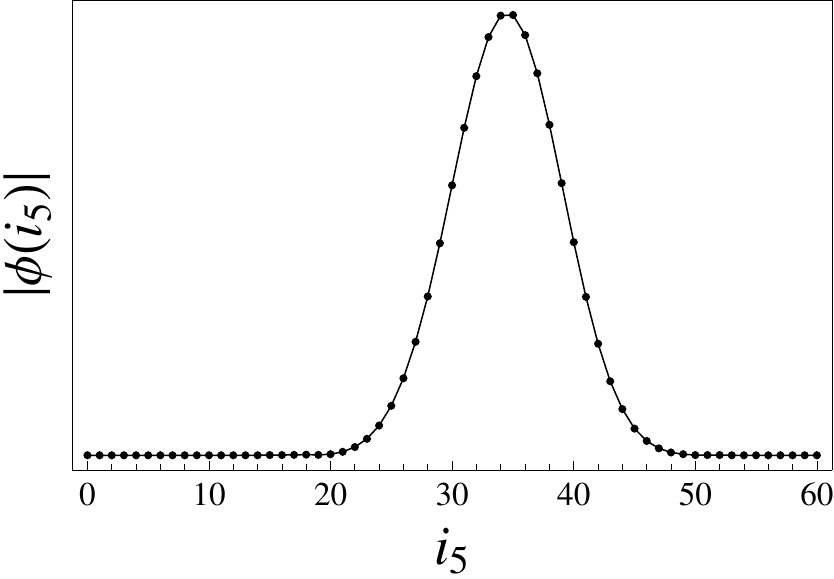}}
\qquad 
\parbox{7.5cm}{\includegraphics[width=7cm]{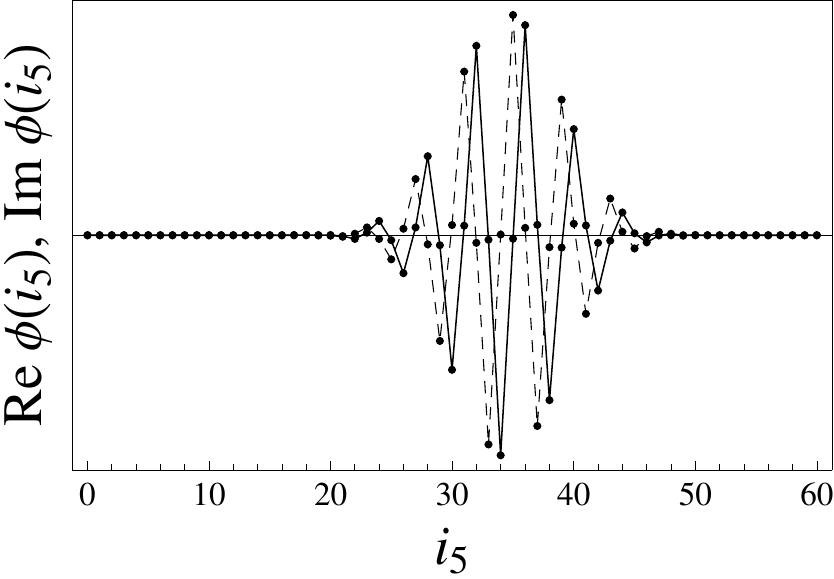}}
\caption{On the left: modulus of the evolved state for the 4-to-1 propagation performed by the flipped vertex ($j_0=30$). On the right: its real and imaginary (dashed) part.}
\label{fig4to1}
\end{figure}
In Fig. \ref{fig4to1} the result of the 4-to-1 evolution for $j_0=30$ is reported. From the plot on the left (the modulus) we can see that the evolved state is a Gaussian peaked on $i_0$ with the same width of the ``incoming" Gaussians. On the right the real and imaginary parts are plotted, and it is clearly visible that the frequency of oscillation is $-\pi/2$, which is exactly the phase opposite to the one of initial packets.\\\noindent In Fig. \ref{fig3to2} are shown the results of the 3-to-2 propagation (moduli), from $j_0=10$ to $j_0=32$ for even $j_0$'s. Compared with the 4-to-1 case, here the Gaussian shape seems not to be conserved, but the state is nevertheless peaked on $i_0$ and presents a $-\pi/2$ phase in both variables; actually a convergence to an elliptic Gaussian is taking place (we explored up to $j_0=56$). Non-Gaussianity has to be imputed to quantum effects. Small deviations from Gaussianity are present also in the 4-to-1-evolution, though less pronounced. Both in the 4-to-1 and 3-to-2 evolution, non-Gaussianity gives rise to deviations of physical expectation values from the classical behavior, well visible in the plots in Fig. \ref{figmedie}.\noindent\\
The physical expectation values (Fig. \ref{figmedie}) are in complete agreement with the expected ones. The small deviations from the semiclassical values gradually disappear as $j_0$ increases. 
\begin{figure}[h]
\centering 
\parbox{7.5cm}{\includegraphics[width=7.5cm]{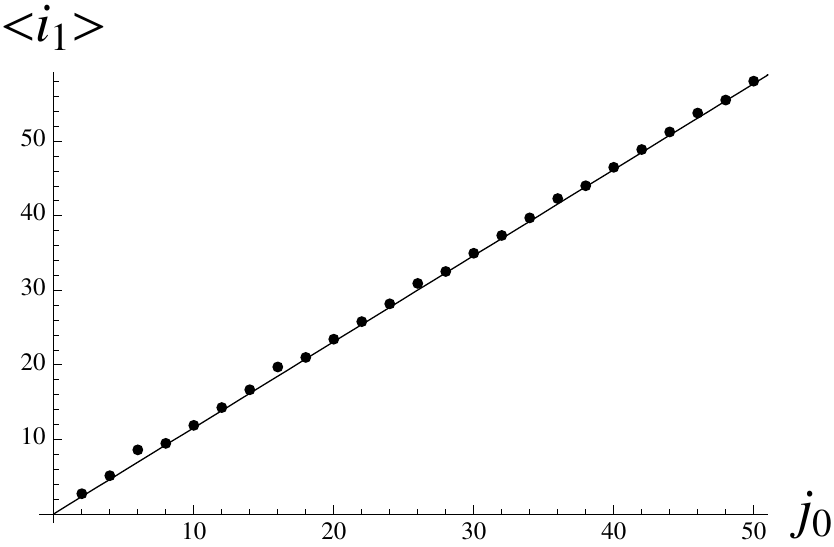}}
\qquad 
\parbox{7.5cm}{\includegraphics[width=7.5cm]{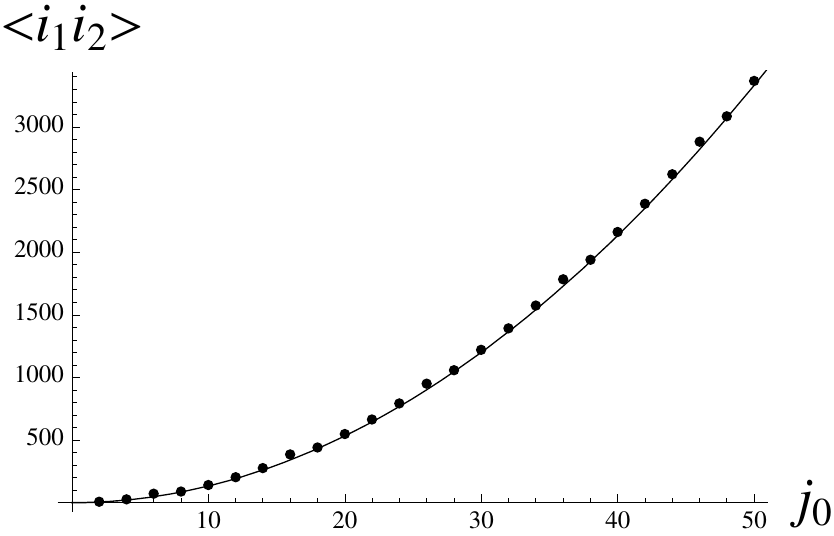}}
\caption{On the left: physical expectation value of $i_1$. On the right: physical expectation value of $i_1 i_2$. The solid line is the expected behavior.}
\label{figmedie}
\end{figure}

\section{Correlation function}
\begin{figure}[h!]
 \centering
 \includegraphics[width=1.5cm]{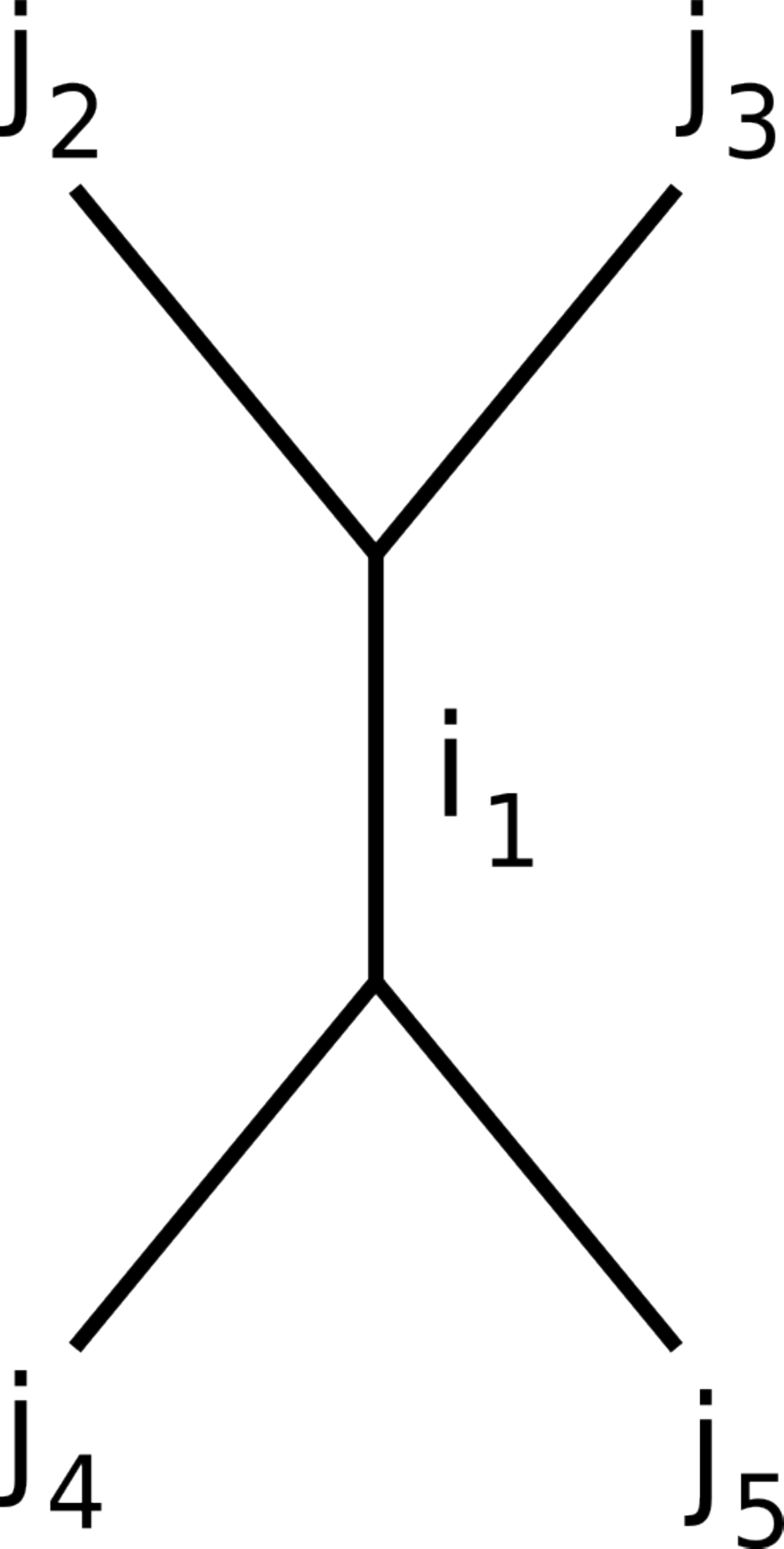}
 \caption{node}
 \label{fignodo}
\end{figure}

\begin{figure}
\centering 
\parbox{7.5cm}{\includegraphics[width=7.5cm]{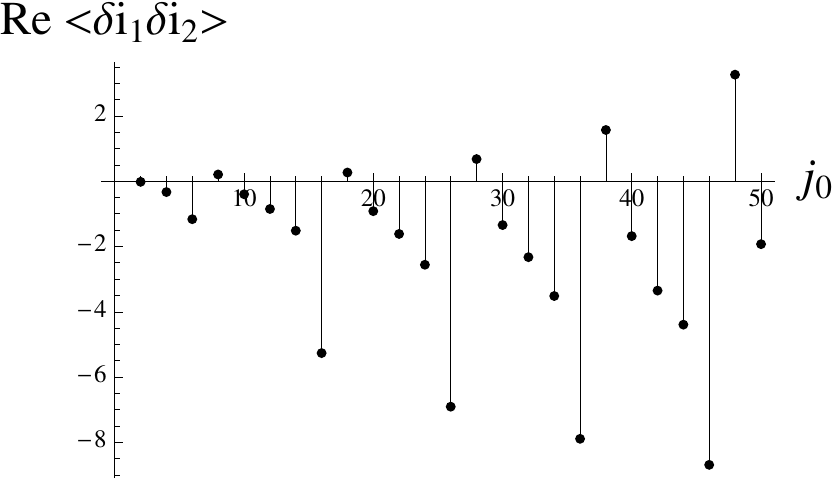}}
\qquad 
\parbox{7.5cm}{\includegraphics[width=7.5cm]{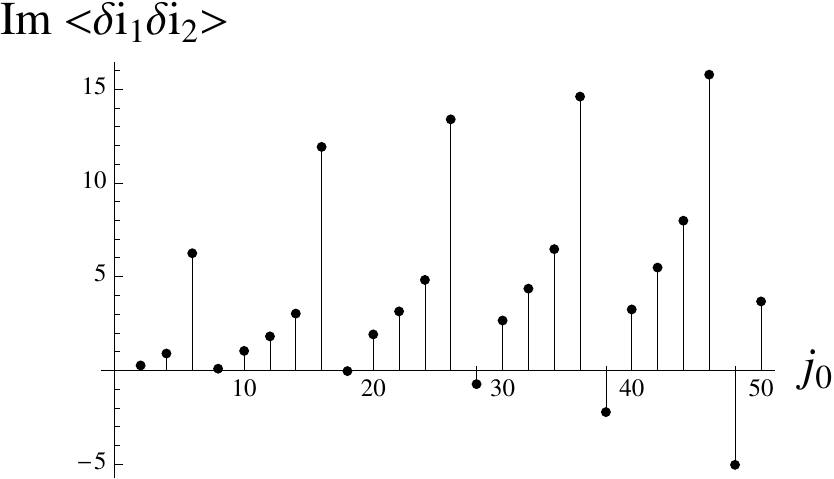}}
\caption{On the left: real part of the intertwiner correlation function, divided by $j_0^2$. On the right: its imaginary part.}
\label{correlation}
\end{figure}

\begin{figure}[h]
\centering 
\parbox{5.2cm}{\includegraphics[width=5.2cm]{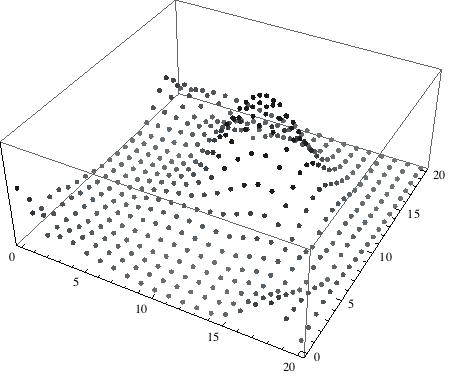}}
\parbox{5.2cm}{\includegraphics[width=5.2cm]{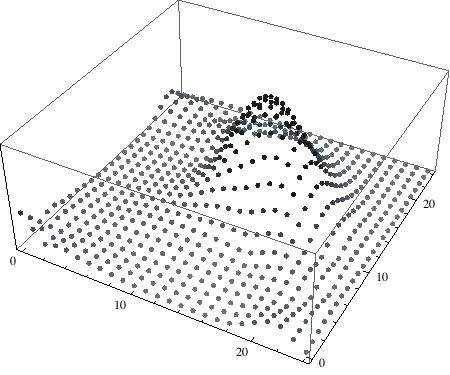}}
\parbox{5.2cm}{\includegraphics[width=5.2cm]{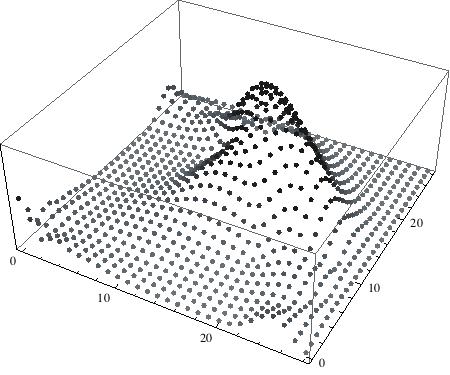}}
\parbox{5.2cm}{\includegraphics[width=5.2cm]{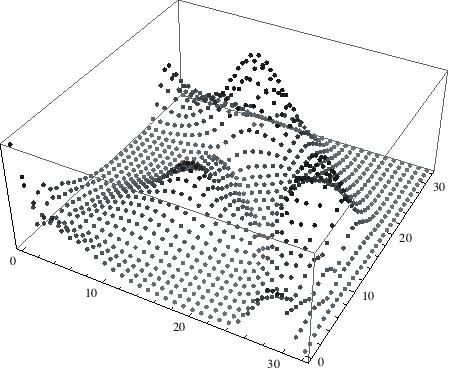}}
\parbox{5.2cm}{\includegraphics[width=5.2cm]{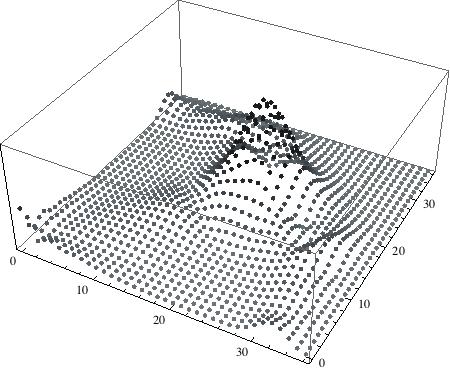}}
\parbox{5.2cm}{\includegraphics[width=5.2cm]{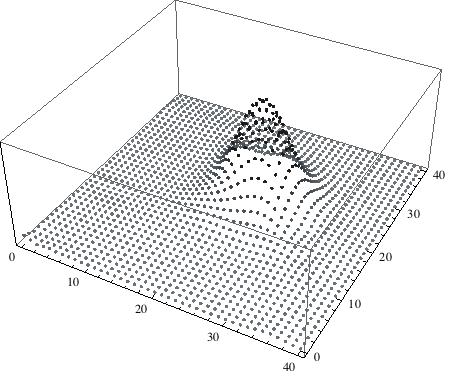}}
\parbox{5.2cm}{\includegraphics[width=5.2cm]{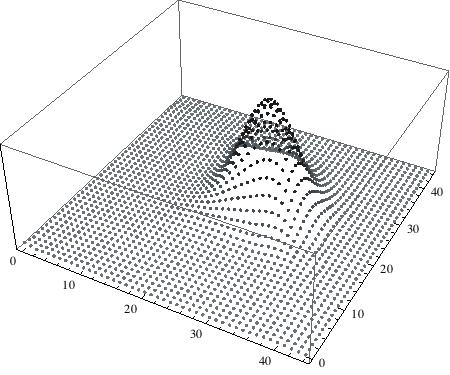}}
\parbox{5.2cm}{\includegraphics[width=5.2cm]{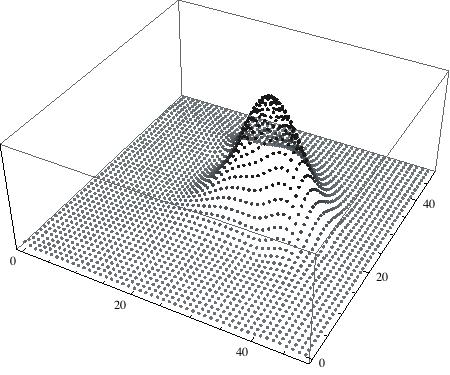}}
\parbox{5.2cm}{\includegraphics[width=5.2cm]{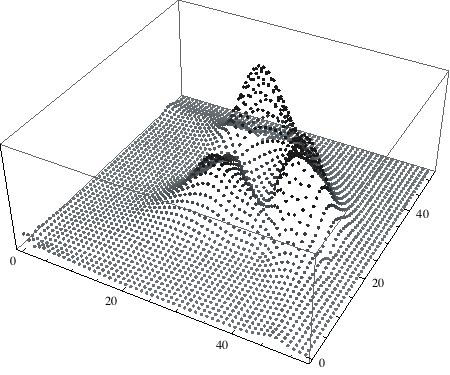}}
\parbox{5.2cm}{\includegraphics[width=5.2cm]{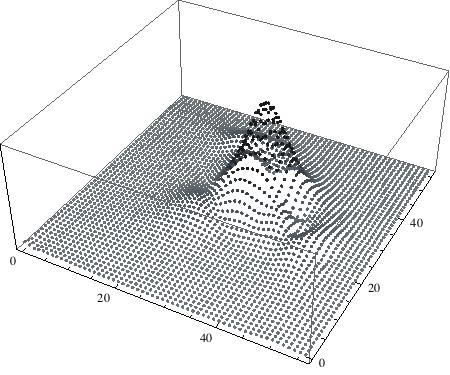}}
\parbox{5.2cm}{\includegraphics[width=5.2cm]{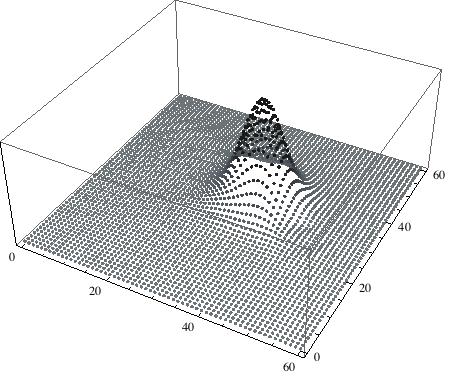}}
\parbox{5.2cm}{\includegraphics[width=5.2cm]{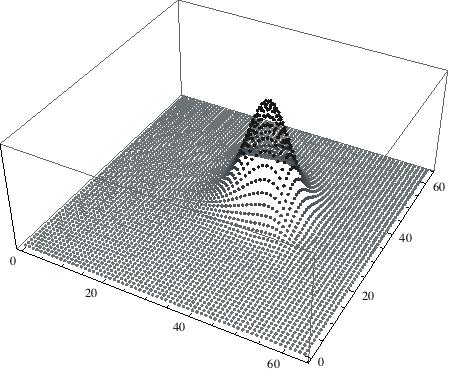}}
\caption{Modulus of the evolved state for the 3-to-2 intertwiner propagation, from $j_0=10$ to $j_0=32$ with step 2.}
\label{fig3to2}
\end{figure}

Here we present some very preliminary results about the graviton propagator in the EPR spinfoam model, in the rough approximation of fixed spins. In other words, we consider the 2-point function over the boundary state \eqref{boundary}. With our approximation, the spin-spin and spin-intertwiner correlation functions vanish, but some components of the graviton propagator are proportional to the intertwiner-intertwiner correlation function and we will study these ones. The 2-point function, or propagator, over the boundary of a 4-simplex is defined by:
\begin{equation}
\label{2point}
G_{mn}^{abcd}=\frac{\sum_{\{j\},\{i\}}W(\{j\},\{i\})(E^{(a)}_m\cdot E^{(b)}_m-n^{(a)}_m\cdot  n^{(b)}_m)(E^{(c)}_n\cdot  E^{(d)}_n-n^{(c)}_n\cdot  n^{(d)}_n)\Psi(\{j\},\{i\})}
{\sum_{\{j\},\{i\}}W(\{j\},\{i\})\Psi(\{j\},\{i\})}
\end{equation}
where $m,n$ and $a,b,c,d$ run over $\{1,\ldots,5\}$, $E_m^a$ is the electric field (densitized triad) operator at the node $m$, projected along the normal $n^{(a)}_m$ in $m$ to the face shared by the tetrahedra $m$ and $a$. In the diagonal-diagonal components ($a=b$, $c=d$) the electric fields act as area operators, so that the 2-point function is essentially an expectation value of two spin insertions ``$\delta j\delta j$", which in our case of fixed spins vanishes. In the diagonal-nondiagonal components ($a=b$, $c\neq d$) the first couple of electric fields gives a spin insertion ``$\delta j$", while the second couple acts nontrivially (in fact the nondiagonal action is the one that ``reads" the intertwiner quantum numbers at nodes) but also those components vanish because of the presence of the spin insertion. The only surviving components are the nondiagonal-nondiagonal; they are quite complicated but some of them are simpler, and we will consider only them. Consider in the boundary state a node $m$ labeled by the virtual spin $i_m$ in a certain pairing, and concentrate only on those $(a,b)$ which are coupled to $i_m$. As an example, if we take the node $m=1$ and the surrounding spins are labeled as in Fig. \ref{fignodo},
then we consider only $a,b\in\{2,3\}$ or $\{4,5\}$ ($a\neq b$). Then consider another node $n$ labeled by $i_n$ and indices $c\neq d$ coupled to $i_n$. For those components the action of graviton operators is diagonal and gives insertions of the kind ``$(\frac{2}{\sqrt 3}\delta i-\delta j-\delta j)(\frac{2}{\sqrt 3}\delta i-\delta j-\delta j)$", so
\begin{align}
\label{2pointourcase}
G_{mn}^{abcd}\propto\frac{\sum_{\{j\},\{i\}} W(\{j\},\{i\})\,\delta i_m\delta i_n\Psi(\{j\},\{i\})}
{j_0^2\sum_{\{j\},\{i\}}W(\{j\},\{i\})\Psi(\{j\},\{i\})}\equiv\frac{\langle\delta i_m\delta i_n\rangle}{j_0^2}.
\end{align}
If the propagator has the Newtonian scaling in the semiclassical regime, it should scale asymptotically as the inverse of $j_0$; equivalently, the quantity $\langle\delta i_m\delta i_n\rangle$ should scale linearly with $j_0$. The plot of $\langle\delta i_m\delta i_n\rangle$ from $j_0=2$ to $j_0=50$ (with step $2$) is shown in Fig. \ref{correlation}. The scaling is clearly not the Newtonian one, and this could be due to our choice of boundary state, which freezes the spins, or maybe to some pathology of the model. The auspicious results about the evolution of wave packets and the physical expectation values seem to exclude the latter possibility.
\section{Conclusions and outlook}
We studied the propagation of semiclassical intertwiners over a 4-simplex, using the EPR spinfoam model (flipped vertex). This approach, introduced in \cite{numerical}, which is a way to study the semiclassical limit of spinfoam models for quantum gravity, turned out to be viable both analytically and numerically, and gave encouraging answers. In particular, certain coherent states turned out to evolve in accordance with classical general relativity. Then we read the results as physical expectation values of observables. In the end, we showed a numerical calculation of the intertwiner correlation function, but the scaling law w.r.t. distance is not the one giving rise to Newton law in the semiclassical regime. This has to be expected, as the approximation of freezing the spins could also prevent intertwiner fluctuations (remember that a classical 4-simplex is fully determined by the ten edge lengths, so if those lengths are given then the dihedral angles between triangles are automatically determined). Though positive, we regard our results as partial and tentative: one should get rid of the ``fixed spin" approximation and see if the wavepacket propagation is still correct and then compute the 2-point function in the semiclassical limit and see if the scaling with distance is the Newtonian one. Further numerical  \cite{Igor} and analytical investigations have already started and we expect in the next future to be able to give more precise answers.
\section*{Acknowledgements} We would like to thank Carlo Rovelli for discussions and Igor Khavkine for correspondence and illuminating programming tricks. E. Alesci and E. Bianchi wish to thank the support by Della Riccia Foundation. 

\end{document}